\documentclass[secnumarabic, graphics,floatfix, nofootinbib,tightenlines,nobibnotes, aps, prl, 12pt]{revtex4}
\usepackage{graphicx}
\usepackage{amsmath}

\begin{document}

\title{A closer look at the influence of
tubular initial conditions on two-particle correlations}
\author{R.P.G.Andrade, F.Grassi, Y.Hama, and \underline{W.-L.Qian}}
\affiliation{Instituto de F\'{\i}sica, Universidade de S\~ao Paulo, SP, Brazil}
\date{Oct. 2009}
\begin{abstract}
In a recent paper, the hydrodynamic code NEXSPheRIO was used in conjunction
with STAR analysis methods to study two-particle correlations as function of $\Delta\eta$ and
$\Delta\phi$. The various structures observed in data were reproduced.
In this work, we discuss the origin of these structures as well as present new
results.
\end{abstract}

\maketitle

\section{I. Introduction}

One of the most striking results in relativistic heavy ion collisions is the 
existence of some structures 
%of near side  and away side ridges
in the two-particle correlations \cite{ridge1,ridge2,dhump1}.
%,dhump2,dhump3}.
One structure has
a long (pseudo)rapidity extent \cite{phobos} and a narrow 
azimuthal extent. 
The other may have a long (pseudo)rapidity extent and has a single or double hump in azimuth.
In order that two-particles A and B emitted at some proper
time $\tau_{f.out}$ appear as correlated, the process that correlated
them must have occurred\cite{lml1,lml2} at some proper time 
$\tau\leq \tau_{f.out} \exp (- |y_A-y_B|/2)$. 
Therefore, the existence of  long range (pseudo)rapidity correlations 
must be related to {\em early times} in the nuclear collisions.

These two-particle azimuthal correlations data have motivated many theoretical 
investigations 
(for a short critical review see e.g.\cite{nagle}).
In many of these approaches, the mechanism is 
closely related to jet
quenching and the response of the medium to the deposited energy.
However, 1) the ridge structure is also seen at low transverse 
momentum\cite{ridge3,ridge4};
2) recent experimental data  seem to indicate no correlation between
high pt trigger and associated ridge particles \cite{ridge6}.
In another class of models, it is suggested that the combined effect of
a longitudinal structure in the initial conditions (IC) and 
transverse expansion is 
responsible for the ridge \cite{voloshin,lml1,lml2}. In this line,
we have
studied the two-particle correlation by using a hydrodynamic code NEXSPheRIO\cite{jun}. 
Both the near-side and double-hump away-side structures were reproduced. In this work,  we discuss how exactly 
these structures appear as well as present new results.

%Recently, the observations of the ridge like structure on near-side\cite{ridge1,ridge2}, and the double hump structure\cite{dhump1,dhump2,dhump3}
%on away side in two-particle correlations have aroused much interest, therefore motivated many theoretical investigations[cite].
%The ridge features a long range pseudorapidity ($\Delta\eta$) correlation in the same direction of a usually high momentum hadron, 
%known as the trigger particle.
%The correlation extends over many units in pseudo-rapidity and relatively narrow in azimuthal angle ($\Delta\phi$).
%In most of theoretical approaches on the market, the mechanism is closely related to jet
%quenching and the response of the medium to this deposited energy.
%However, recent experimental analysis concludes that there seems no jet-ridge "cross talk"[cite].
%On the opposite side of the trigger particle, one also finds interesting physics.
%Di-hadron azimuthal correlations revealed a double hump structure on the away side. 
%It has been postulated that this structure is due to a Mach-cone produced by a super-sonic parton[cite].

\section{II. Ridges and peak in NeXSPheRIO}

\begin{figure}[!htb]
\vspace*{-0.0cm}
\includegraphics[width=12.cm]{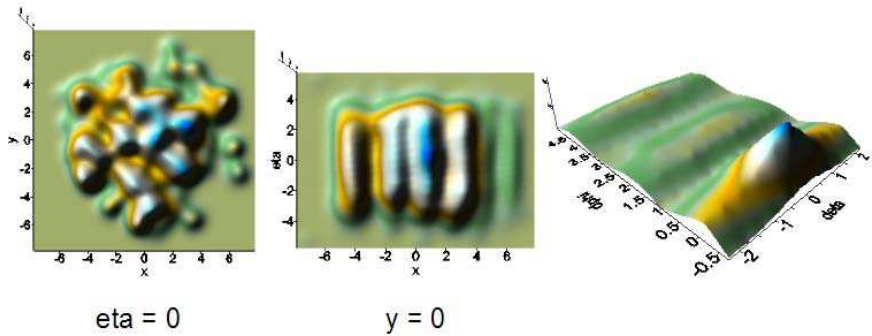}
\vspace*{-0.0cm}
\caption{NeXSPheRIO initial energy density in the transverse (left)
and reaction (center) planes for a central Au+Au collision at 200 GeV A.
Two-particle correlation (right).}
 \label{figIC1}
\end{figure}

\begin{figure}[!htb]
\vspace*{-0.0cm}
\includegraphics[width=6.cm]{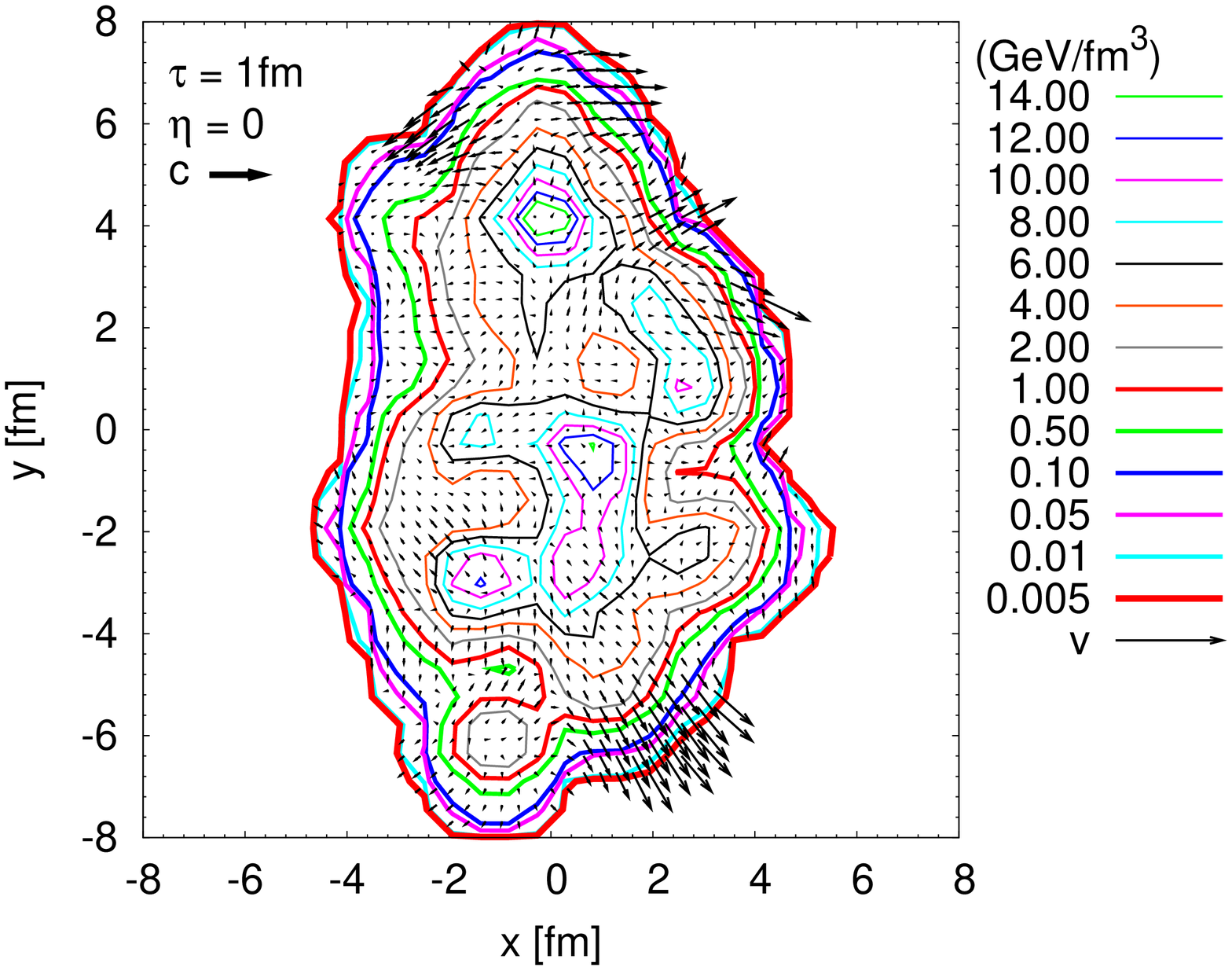}
\includegraphics[width=6.cm]{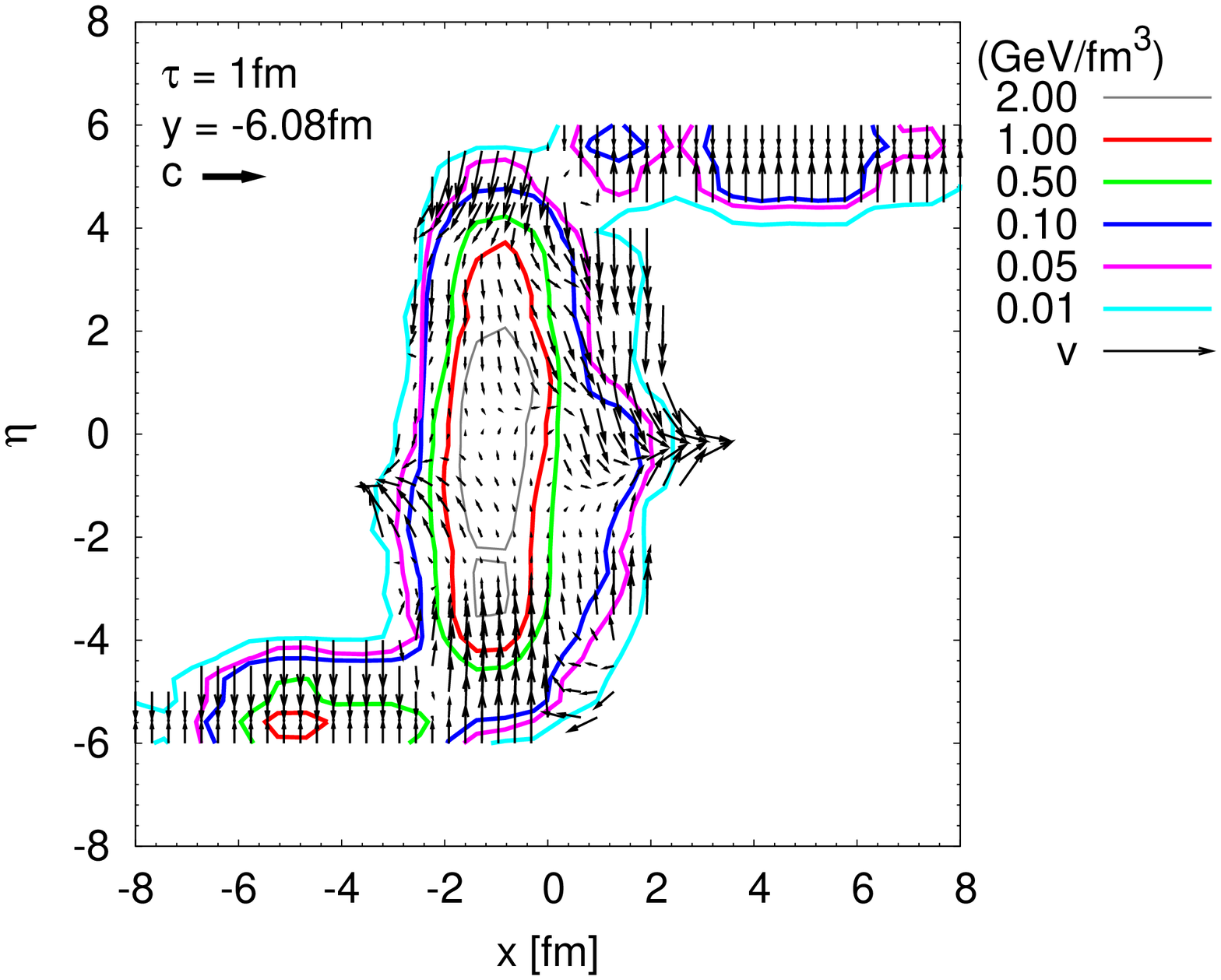}
\vspace*{-0.0cm}
\caption{NeXSPheRIO initial fluid velocity for a Au+Au collision at 200 GeV A
in the 25-35\% centrality window.} 
\label{figIC2}
\end{figure}

The NeXSPheRIO code uses IC from the microscopic code NeXus \cite{nexus} 
and solves the 
hydrodynamic equations with the
SPheRIO code \cite{topics},  on an event by event basis.
%The SPheRIO code
% is an implementation of the entropy representation of the Smoothed Particle Hydrodynamics (SPH) model 
%for relativistic high-energy collisions, and it has been investigated and developed within 
%the S\~{a}o Paulo - Rio de Janeiro Collaboration. The
%SPH method was first introduced for astrophysical applications\cite{astro}, 
%and later adapted for relativistic heavy-ion collisions\cite{va}.
%It parameterizes the matter flow in terms of discrete Lagrangian coordinates, called SPH particles. 
%In terms of SPH degrees of freedom, the equations of motion can be derived by using the variational principle\cite{vp}. 
%The main advantage of the method is that it is rather robust to deal with any kind of geometrical structure and violent dynamics. 
%For example, shock wave phenomena can be treated without numerical difficulty, provided the size of SPH particles is appropriately chosen. 
%SPheRIO code can be connected to microscopic IC event generators (in the present version NeXus)
% as NEXUS[cite] or EPOS[cite] directly,
%in such a way that  calculations can be carried out on event by event basis.
Figure \ref{figIC1} shows a typical example of IC with tubular structures along the 
collision axis. In our model, this is what causes the 
long (pseudo)rapidity extent of the two-particle correlations.
We note that in figure \ref{figIC1}, the near-side structure is composed of a ridge and
a peak, as seen experimentally. 
In our approach, the IC constructed by ``thermalizing'' NeXus 
output\cite{topics} do not explicitly involve jets, 
these are however not totally forgotten in the IC as they leave some 
localized region with higher transverse fluid velocity as shown in figure 
\ref{figIC2}. This region is not correlated with a tube so peak and ridge are independent.

We also note in figure \ref{figIC1} the existence not only of the 
near-side ridge but also of the double-hump away-side ridge. This result was 
obtained in \cite{jun} using the same methods as in the STAR analysis, in particular elliptic flow was removed using ZYAM.
In this work, a totally different analysis is done, in particular
the directions of the different event planes 
(which is calculable in our model\cite{v2}) are aligned,
then the mixed event contribution is obtained and subtracted from the raw 
correlation.
By doing this,  the effects of  $\eta$ distribution shape and flow are removed in a single 
step.

%The model has also been employed to study the centrality and momentum threshold dependence of the two-particle correlation
%as shown in Fig.1-2, which are in qualitatively agreement with the experimental data.
%In Fig.3 we studied in-plane and out-of-plane effect of two-particle correlation, as one can see clearly, 
%the double hump structure becomes more pronounced
%as the angle of trigger paticle increase, it peaks at $\phi_{s} \sim 45^{\circ}$ and decreases again when it
%goes to out-of-plane direction, showing good agreement with experimental data[cite].
%In the above calculations, we first aligned the directions of event planes of different events 
%since it is calculable in our model\cite{v2},
%then the mixed event contribution is calculated and subtracted from the raw 
%correlation.
%In this way the effect of mixed event and of flow is removed in a single step.
The fact that this new analysis leads to similar results as in \cite{jun}
reinforces both approaches. However it is still unclear what causes in 
detail the near and away-side structures as well as what fixes the position
of the double-hump in the away-side ridge.

\section{III. Origin of the ridges in a simplified model}

%In our approach, IC constructed by NEXUS does not explicitly involve jets.
%Therefore, different from most of interpretations on the market, 
%the resulting correlation structure is probably not related to jets,
%which nevertheless coincides with recent experimental observations[cite].
%On the other hand, for exact the same reason, the double peak structure observed on the away side in our model
%cannot be ascribed to any mechanism related to deflected jet[cite] or conical emission[cite].
%One question still remains, that is how does one understand the away side correlation results in the framework of 
%hydrodynamic model.

\begin{figure}[!htb]
\vspace*{-0.0cm}
\includegraphics[width=5.cm]{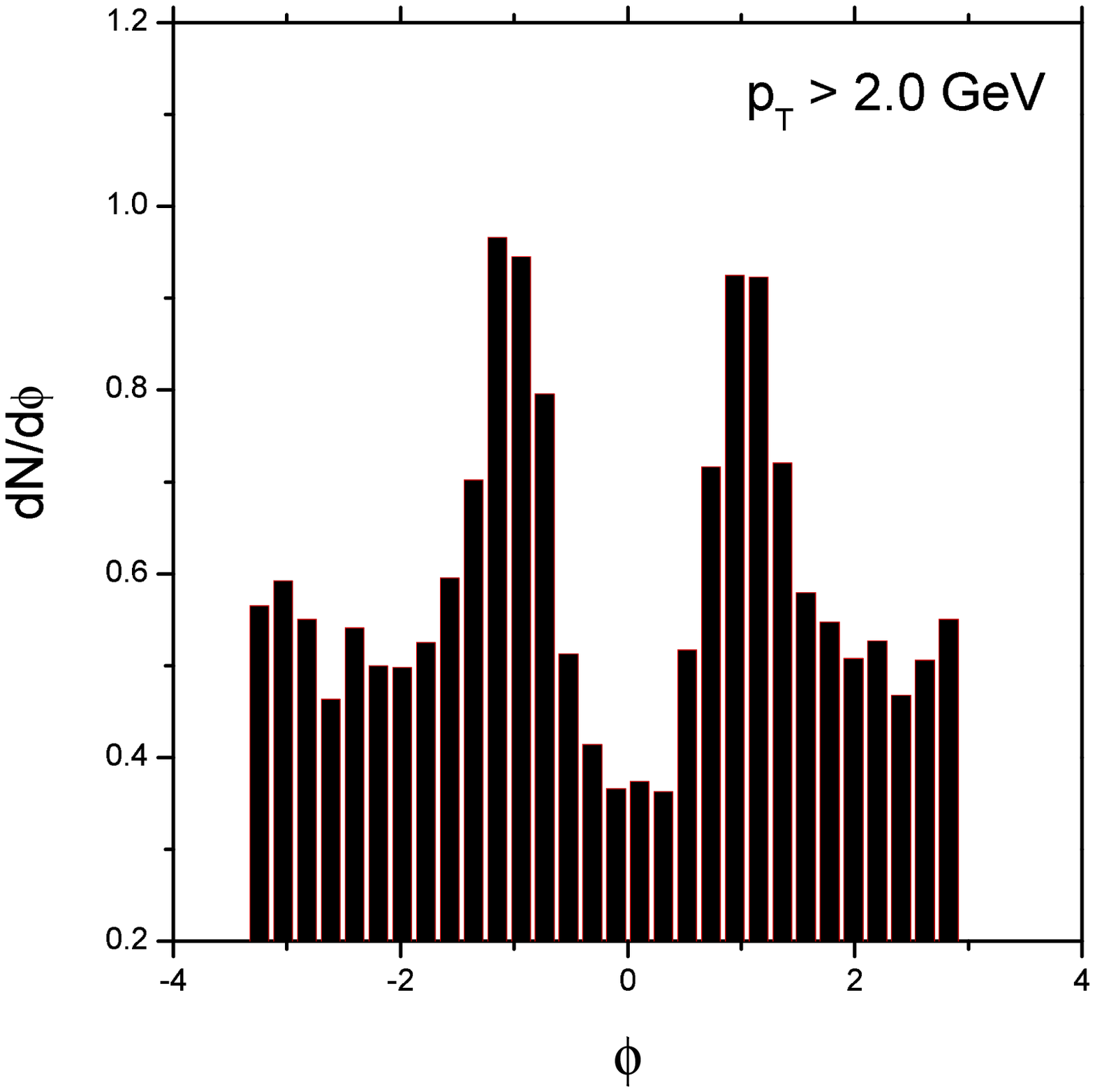}
\includegraphics[width=5.cm]{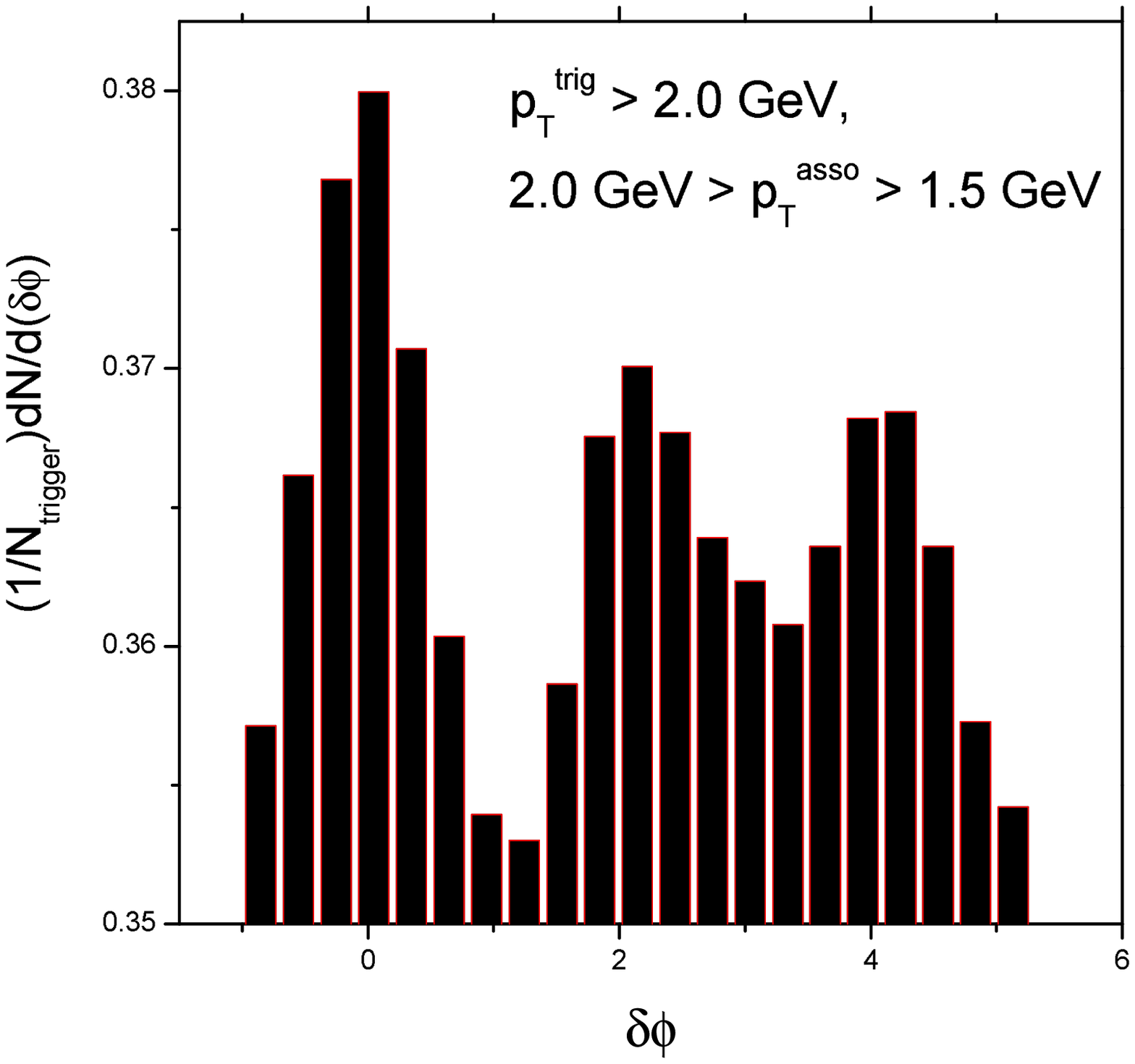}
\vspace*{-2.3cm}
\caption{Single (left) and two (right) particle  angular distribution in the simplified model.} 
\label{angdist1}
\end{figure}

To investigate the origin of the ridges,  we use a simplified two-dimensional
model. This model consists of a slice of 
matter which initially has a high energy density spot in a smooth background. 
This slice subsequently undergoes   transverse expansion and 
boost-invariant longitudinal expansion.
More details are given in
\cite{ismd09}. The single particle angular distribution has not a single peak as one might expect but
two peaks located on both sides of the position of the tube as seen in figure \ref{angdist1} (left).
This double peak structure is observed for all transverse momenta at more 
or less the same position \cite{ismd09} and its location is in agreement

\begin{figure}[!htb]
\vspace*{-0.0cm}
\includegraphics[width=7.cm]{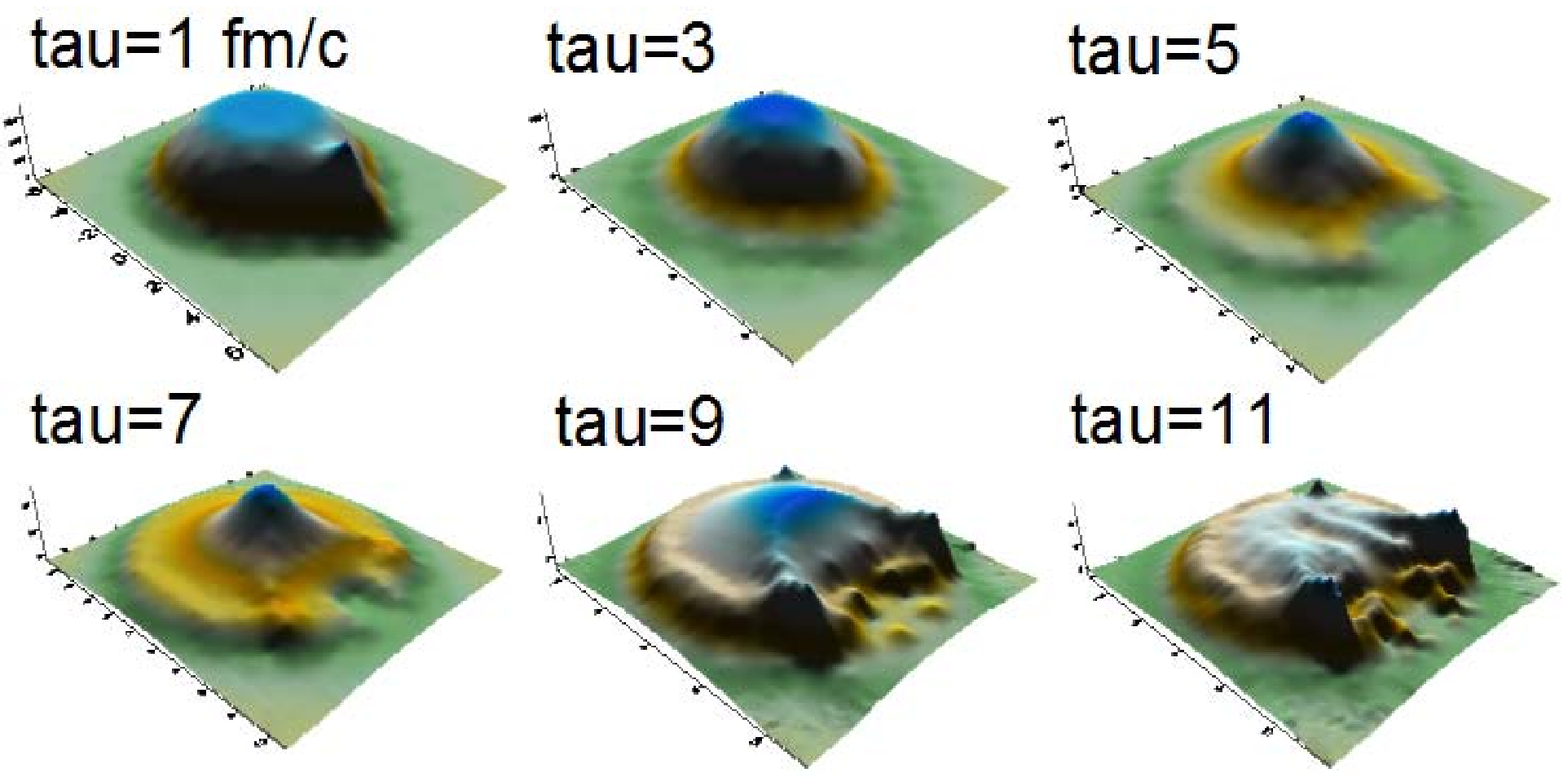}
\includegraphics[width=6.7cm]{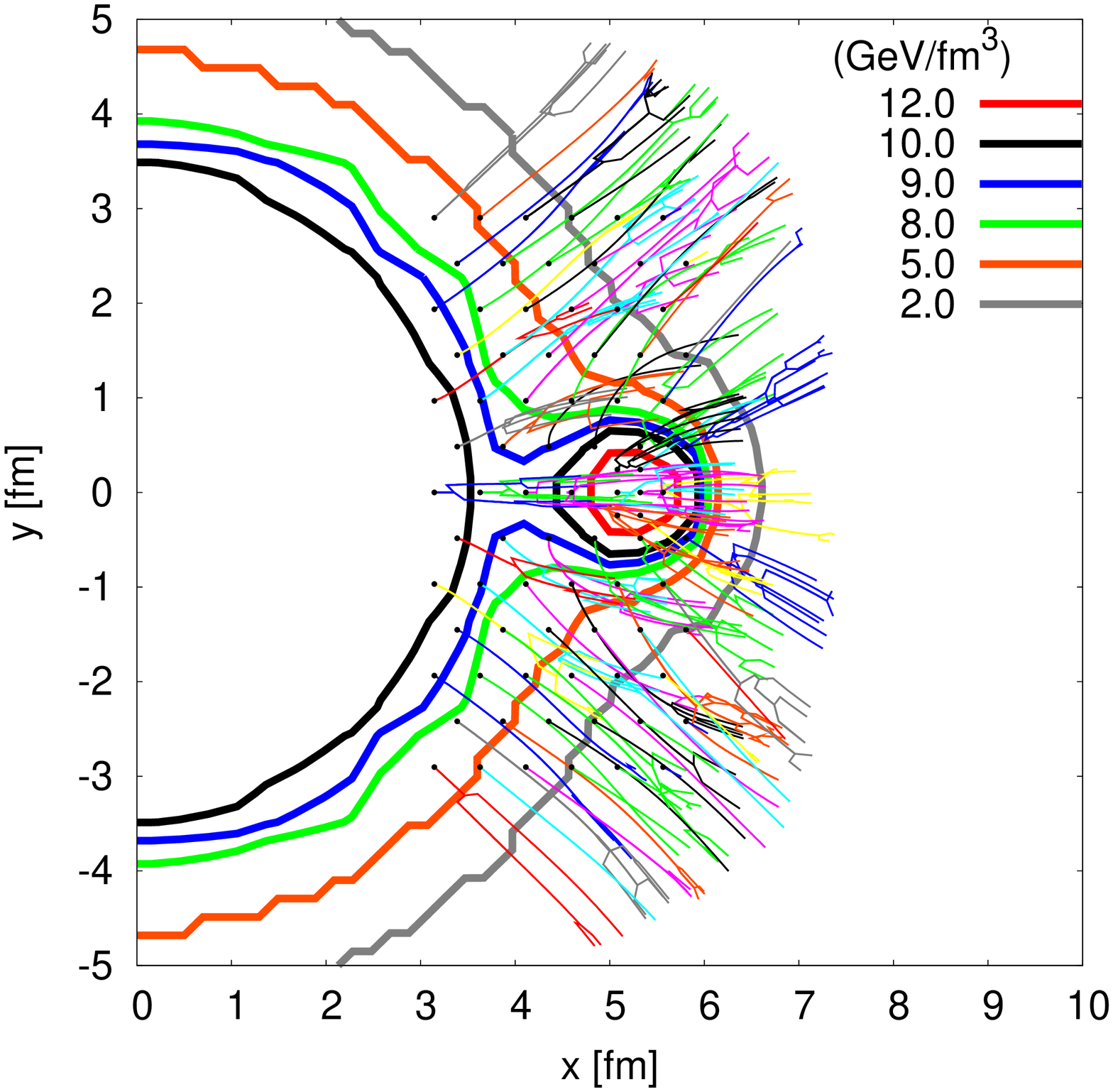}
\vspace*{-0.0cm}
\caption{Temporal evolution of energy density for the simplified model (left).
Trajectories of the fluid cells around the tube (right).} 
\label{hydrotube}
\end{figure}

\noindent with data (for a central collision).
The reason, as seen in figure \ref{hydrotube} is that
the effect of the tube is to 
deflect the otherwise isotropic radial flow.

From figure \ref{angdist1} (left), we can guess how the two-particle angular 
correlation will be. The trigger particle is more likely to be in one of the 
two peaks. We first choose the left-hand side peak. The associated particle is 
more likely to be also in this peak i.e. with $\Delta \phi=0$ or
in the right-hand side peak with $\Delta \phi\sim +2$. If we choose the
trigger particle  in the right-hand side peak, the associated particle is 
more likely to be also in this peak i.e. with $\Delta \phi=0$ or
in the left-hand side peak with $\Delta \phi\sim -2$.
So the final two particle angular 
correlation  must have a large central peak at $\Delta \phi=0$ and two smaller
peaks respectively at $\Delta \phi\sim \pm 2$.
Figure \ref{angdist1} (right) shows that this is indeed the case.
The peak at $\Delta \phi=0$ corresponds to the near-side ridge and the peaks
at $\Delta \phi\sim \pm 2$ form the double-hump ridge. We have checked  that
this structure is robust by studying  the effect of the height and shape of the background, initial velocity, height, radius and location of the tube \cite{ismd09}. 
We are extending this model to non-central collisions. 

\section{IV. Comparison of NeXSPheRIO results with experimental data}

Now that the origin of the ridges  is clarified, we return
to a comparison of NeXSPheRIO results with data.
Since, as already explained, the jets are ``thermalized'' in our model, a

\begin{figure}[!htb]
\vspace*{-0.0cm}
\includegraphics[width=6.cm]{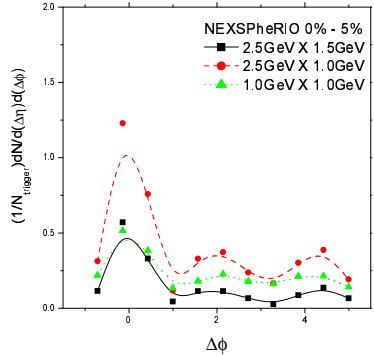}
\vspace*{-3.cm}
\caption{NeXSPheRIO dependence of the ridges on the $p_t$ cutoffs  for 
Au+Au collisions at 200A GeV.} 
\label{ptridge}
\end{figure} 

\noindent precise quantitative comparison at this stage is not possible. Nevertheless several 
qualitative comparisons can be done.

For fixed $p_t^{trig}$ and increasing $p_t^{assoc}$, the near-side and away-side peaks decrease as seen in figure \ref{ptridge} for central collisions
(this is generally expected since the number of associated particles decreases). 
On the other side,  
for fixed $p_t^{assoc}$ and
 increasing $p_t^{trig}$, the peaks increase. 
This behavior is
in agreement with data (fig. 36 in \cite{dhump3}).

\begin{figure}[!htb]
\vspace*{-0.0cm}
\includegraphics[width=8.cm]{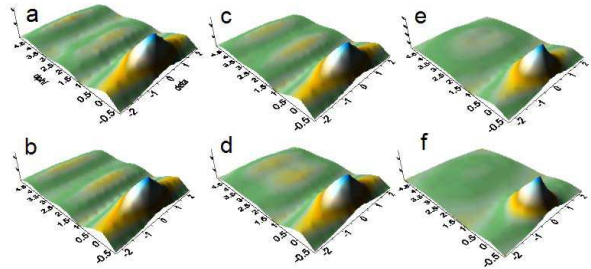}
\vspace*{-0.0cm}
\caption{Two-particle correlation, as computed with our NeXSPheRIO code, for 
different centrality windows of
Au+Au collisions at 200A GeV(a: 0-5\%, b: 5-10\%, c: 10-20\%, d: 20-30\%, e: 
30-40\%, f: 40-50\%).  $p_t^{trig}
 > 2.5$ GeV and $p_t^{assoc}
 > 1.5$ GeV.} 
\label{centrridge}
\end{figure}

When going from central to peripherical collisions, the near-side ridge 
decreases and the away-side ridge changes from double to single hump, as seen in figure \ref{centrridge} and in conformity with data(fig. 36-38 in \cite{dhump3}).

\begin{figure}[!htb]
\vspace*{-0.0cm}
\includegraphics[width=8.cm]{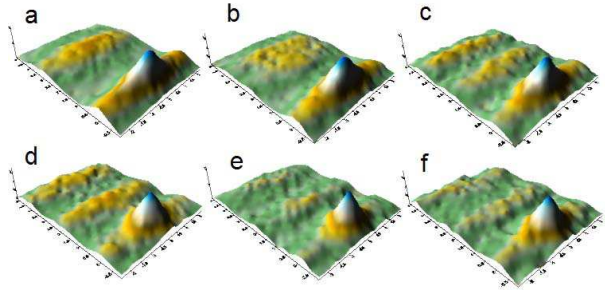}
\vspace*{-0.0cm}
\caption{Two-particle correlation, from NeXSPheRIO, for 20-30\% centrality
Au+Au collisions at 200A GeV, for different $\phi_s$ (a:
 $\phi_s = 0^o-15^o$, b: $\phi_s = 15^o-30^o$,
 c: $\phi_s = 30^o-45^o$,
d:$\phi_s = 45^o-60^o$,
e: $\phi_s = 60^o-75^o$,
f: $\phi_s = 75^o-90^o$.
$p_t^{trig}
 > 3.$ GeV and $p_t^{assoc}
 > 1.$ GeV.} 
\label{inoutridge}
\end{figure}

The correlation can be studied as a function of the trigger particle angle 
with 
relation to the event plane. In figure \ref{inoutridge} for
 a mid-central window, the away-side 
ridge changes from single peak for in-plane trigger to double peak for out-of-plane trigger. For central collisions (not shown), it is always double-peaked.
This is in accordance with data (fig.1 in \cite{inout}).

Some other qualitative features can also be mentioned.
The near-side ridge seems to be present for untriggered correlation 
as found experimentally \cite{ridge3,ridge4} but
our analysis method is different from the experimental one and this result
is still being checked. As explained above, the near-side ridge is independent 
of the jet peak, also in agreement with data \cite{ridge6}. Due to its origin, we
expect the  near-side ridge to have similar composition as the bulk and its spectra
should be a little harder than bulk (see figure \ref{angdist1} left and figgure 5 of \cite{ismd09}), 
also in line with data \cite{ridge5,ridge6}. We expect a similar behavior
for the away-side ridge in accordance with data when available \cite{ridge6,ridgex}.
Finally, for non-central collisions, due to their 
origin, we expect asymmetry in the ridges azimuthal correlation as function of the trigger angle with respect to the event plane; for the near-side ridge this has been observed 
in \cite{ridge7}.

\section{V. Conclusion}% Possible tests of this model}

In conclusion, the hydrodynamic expansion
starting from fluctuating tubular IC 
produces the various structures observed in the two-particle
correlations \cite{jun}. 
We showed in this paper explicitly how this happens, 
using a simplified single tube model: this tube deflects matter in two 
directions which results in the near-side and away-side ridges.
In addition we show that NeXSPheRIO code reproduces several other observed
characteristics of the two-particle correlations.

\begin{figure}[!htb]
\vspace*{-0.0cm}
\includegraphics[width=6.cm]{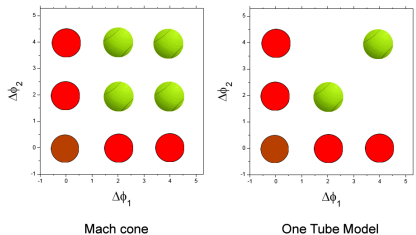}
\vspace*{0.0cm}
\caption{Schematic three particle correlations as function of the angles of
the two associated particles with respect to the trigger particle.}
\label{3dist}
\end{figure}

Other models have been suggested to explain 
some of theses data.
Three particle correlations might provide a way to discriminate: 
some models such as the Mach cone one
predict that the associated particles  emerge at two 
angles leading to figure \ref{3dist} (left) while others such
as the deflected jet model result in particles (if not flying with the trigger) emerging at
one angle giving rise to figure \ref{3dist} (right). Such a figure is also
expected for the simplified 
one tube model and approximately for NeXSPheRIO central collisions.
We are working on
a precise prediction and comparison with data 
\cite{star3p} . 

It seems that an interesting observable would be 
2+1 correlations (first associated particle fixed in a narrow kinematical interval), 
which requires less statistics and leads to a clean prediction. 
For example choosing the trigger from one peak in figure \ref{angdist1} (left)
and the first associated particle from the other peak, 
the second associated particle will be from any of the two peaks.
This means that in two particle correlation between the second associated particle and the trigger 
presented in terms of $\Delta \eta$ and $\Delta \phi$, there would be two 
stripes. In contrast, the Mach-cone model would lead to three stripes.

\section{VI. Acknowledgments}
We thank for discussions with Takeshi Kodama, Jun Takahashi, Bernardo Tavares,
Fuqiang Wang, Guoliang Ma, Paul Sorensen, Klaus Werner. We acknowledge funding from 
Funda\c{c}\~ao de Amparo \`a Pesquisa de Estado de S\~ao Paulo,
FAPESP, and Conselho Nacional de Desenvolvimento Cientit\'{\i}fico e 
Tecnol\'ogico, CNPq.

\end{document}